\documentclass[10pt]{iopart}
\providecommand{\mcal}{\ensuremath{\mathcal}}
\newcommand{\opr}[1]{\ensuremath{\mathbf{\mathsf{#1}}}}
\newcommand{\dom}[1]{\ensuremath{\mathcal{D}(\opr{#1})}}

\newcommand{\nul}[1]{\ensuremath{\mathcal{N}(\opr{#1})}}

\newcommand{\abs}[1]{\ensuremath{\left|#1\right|}}

\begin{document}

\title[Quantum Time of Arrival Problem]{Canonical Pairs, Spatially Confined Motion\\ and the Quantum Time of Arrival Problem}

\author{Eric A. Galapon\footnote[3]{email: egalapon@nip.upd.edu.ph}}

\address{Theoretical Physics Group, National Institute of Physics\\ University of the Philippines, Diliman Quezon City\\ 1101 Philippines}

\begin{abstract}
It has always been believed that no self-adjoint and canonical time of arrival operator can be constructed within the confines of standard quantum mechanics. In this Letter we demonstrate the otherwise. We do so by pointing out that there is no a priori reason in demanding that canonical pairs form a system of imprimitivities. We then proceed to show that  a class of self-adjoint and canonical time of arrival (TOA) operators can be constructed for a spatially confined particle. And then discuss the relationship between the non-self-adjointness of the TOA operator for the unconfined particle and the self-adjointness of the confined one.
\end{abstract}
\pacs{03.65 Bz}
\bigskip
The question of \textsl{when a given particle prepared in some initial quantum state arrive at a given spatial point} is a legitimate quantum mechanical problem requiring more than a parametric treatment of time. In standard quantum formulation, this raises the time of arrival at the level of quantum observable. And at this level the time of arrival (TOA) distribution is supposedly derivable from the spectral resolution of a certain self-adjoint TOA-operator canonically conjugate to the driving Hamiltonian. Thus the question of \textsl{when} translates to the question of \textsl{what is the TOA-operator}. But can one construct such an operator? The consensus is a resounding \textsl{no}. This consensus goes back to the well known Pauli's theorem which asserts that the existence of a self-adjoint time operator (of any kind) implies that the spectrum of the Hamiltonian is the entire real line contrary to the generally discreet and semibounded Hamiltonian operator \cite{pauli}. The embargo imposed by Pauli's theorem has led to various treatments of the problem within and beyond the usual formulation of quantum mechanics \cite{works,trans,freetime,egus,grot}. 

However, we have recently shown---following Pauli's own method of proof---the consistency of a bounded, self-adjoint time operator canonically conjugate to a Hamiltonian with a non-empty point spectrum, discreet or semibounded \cite{galapon}. This denies the sweeping generalization of Pauli's conclusion. Motivated by this development, we pose the question \textsl{If Pauli's well known argument can not be correct, then why there is a prevalent failure in constructing a self-adjoint and canonical TOA-operator?} In this Letter we attempt to answer this question specifically for the free particle in one dimension within the confines of the standard single Hilbert space quantum mechanics.  We approach the question by first addressing the issue of quantum canonical pairs at the foundational level. Using the insight we gain in clarifying canonical pairs, we proceed in investigating the TOA-problem under the assumption that the particle is confined. We show that self-adjoint and canonical TOA operators for the spatially confined particle can be constructed. We then discuss the relationship between the self-adjointness of the TOA operator in a bounded space and the non-self-adjointness of the same operator in unbounded space.  

We first address the issue of quantum canonical pairs. A major impediment in constructing a self-adjoint time of arrival operator has been the conviction among workers  that a given pair of self-adjoint operators, $\opr{Q}$ and $\opr{P}$, satisfying the canonical commutation relation $(\opr{QP}-\opr{PQ})\subset i\hbar\,\opr{I}$ (where $\opr{I}$ is the identity operator of the Hilbert space) yield a transitive system of imprimitivities over the entire real line \cite{trans}. That is if $\Delta$ is a Borel subset of $\Re$, and $\Delta \rightarrow \opr{E_Q}(\Delta)$ and $\Delta \rightarrow \opr{E_P}(\Delta)$ are the respective projection valued measures of $\opr{Q}$ and $\opr{P}$, then for every $\alpha,\beta\in\Re$
\begin{eqnarray}
	U_{\alpha}^{-1}\opr{E_Q(\Delta)}U_{\alpha}&=\opr{E_Q(\Delta_{\alpha})}, 				\label{com1}\\
	V_{\beta}^{-1}\opr{E_P(\Delta)}V_{\beta}&=\opr{E_P(\Delta_{\beta})}, 				\label{com2} 
\end{eqnarray}

\noindent where $U_{\alpha}=\exp(i\alpha \opr{P})$ ($V_{\beta}=\exp(i \beta \opr{Q})$), and $\Delta_{\alpha}=\{\lambda:\lambda-\alpha\in\Delta\}$ ($\Delta_{\beta}=\{\lambda:\lambda-\beta\in\Delta$\}). Equations (\ref{com1}) and (\ref{com2}) imply that the spectrum of $\opr{Q}$ and $\opr{P}$  is the entire real line. This automatically forbids the construction of a self-adjoint time operator canonically conjugate to a given semibounded or discrete Hamiltonian if  equations (\ref{com1}) and (\ref{com2}) are imposed upon every physically acceptable canonical pair. 

However, the above conviction can be traced either from Pauli's theorem or from analogy to the properties of the position,$\opr{q}$, and momentum, $\opr{p}$, operators in unbounded free space (for example ref.\cite{gott}). Having addressed Pauli's objections \cite{galapon}, we point out that the analogy is false. It is well known that the pair $(\opr{q},\opr{p})$ satisfy the canonical commutation relation and equations (\ref{com1}) and (\ref{com2}). However, it is not so well known that they satisfy (\ref{com1}) and (\ref{com2}) do not follow from them satisfying the canonical commutation relation. In fact it is the converse. That $(\opr{q},\opr{p})$ satisfy $(\opr{qp}-\opr{pq})\subset i\hbar\,\opr{I}$ follows from the fundamental axiom of quantum mechanics that the propositions for the location of an elementary particle in different volume elements are compatible, and from the fundamental homogeneity of free space, i.e. points in $\Re$ are indistinguishable \cite{macjau}. The former naturally leads to the self-adjoint position operator $\opr{q}$ in $\Re$; while the later requires the existence of a unitary operator generated by the momentum operator such that the PV measure of $\opr{q}$ satisfy equation (\ref{com1}). Then symmetry dictates that the PV measure of $\opr{p}$ must satisfy equation (\ref{com2}). Now equation (\ref{com1}) and the fact that $U_{\alpha}$  and $V_{\beta}$ form a representation of the additive group of real numbers lead to the well known Weyl's commutation relation, $U_{\alpha}V_{\beta}=e^{i\alpha\beta}V_{\beta}U_{\alpha}$. This relation finally implies the canonical commutation relation $(\opr{qp}-\opr{pq})\subset i\hbar\,\opr{I}$ enjoyed by $\opr{q}$ and $\opr{p}$. For a free particle in a box, the points in the spatial space available to the particle are distinguishable, the walls being the distingushing factor, e.g. one point can be nearer to the left wall than another point. The bounded space for the particle then is not homogenous and equation (\ref{com1}) can not be imposed upon the position operator. Doing so is imposing homogeneity in an intrinsicaly inhomegenous space. However, it is known that the self-adjoint position and momentum operators for the trapped particle satisfy the canonical commutation relation without satisfying equations (\ref{com1}) and (\ref{com2}). It should then be clear that $\opr{q}$ and $\opr{p}$ (in $\Re$) satisfy equations (\ref{com1}) and (\ref{com2}) not because they are canonically conjguate but because of an underlying quantum mechanical axiom and a fundamental property of free unboundned space. And that they are canonically conjugate because of these two.

Therefore we are led to redefine and reinterpret quantum canonical pairs. We expand the class of physically acceptable canonical pairs to include any given pair of densely defined, self-adjoint operators,(\opr{Q},\opr{P}), in a separable, infinite dimensional Hilbert space, $\mcal{H}$, satisfying the canonical commutation relation in some nontrivial, proper (dense or closed)  subspace $\mcal{D}_c\subset \mcal{H}$; that is, $(\opr{QP}-\opr{PQ})\varphi=i\hbar\varphi$ for all $\varphi\in\mcal{D}_c$.  (The known fact that there are numerous non-unitarily equivalent solutions to the canonical commutation relation \cite{galapon,nelson} assures us of the richness of canonical pairs beyond those satisfying (\ref{com1}) and (\ref{com2}).) That a given pair is canonical in some sense---e.g. the pair satisfies equations (\ref{com1}) and (\ref{com2}), or one of the pair is bounded and thus do not satisfy (\ref{com1}) and (\ref{com2})---is consequent to a set of underlying fundamental properties of the system under consideration or to the basic definitions of the operators involved or to some fundamental axioms of the theory. It is concievable to impose that a given pair be canonical as a priori requirement based, say, from its classical counterpart, but not the sense the pair is canonical without a deeper insight, say,  into the underlying properties of the system. In other words, we don't impose in what sense a pair is canonical if we don't know much, we derive in what sense instead. Furthermore, we claim that if a given pair is known to be canonical in some sense, then we can learn more about the system or the pair by studying the structure of the sense the pair is canonical. 

Having cleared our way through equations (\ref{com1}) and (\ref{com2}), now we can take another look at the free time of arrival problem. Classically if the position of a given particle in one dimension is $q$ and its momentum is $p$, its time of arrival at the origin is given by $T=-\mu qp^{-1}$ where $\mu$ is the mass of the particle; $T$ is canonically conjugate to the free Hamiltonian $H=(2\mu)^{-1}p^2$, i.e. $\{H,T\}=1$. In the course of history of the quantum time of arrival problem (and related problems), $T$ has served as the starting point in numerous attempts in constructing time of arrival operator.  Various quantization schemes lead to the totally symmetric quantized form of $T$,
\begin{equation} \label{qtoa}
	\opr{T}=-\mu \frac{\opr{qp^{-1}+p^{-1}q}}{2}.
\end{equation}
\noindent in which $\opr{T}$, $\opr{q}$ and $\opr{p}$ are the operator versions of $T$, $q$ and $p$, respectively. Formally (\ref{qtoa}) is canonically conjugate to the free Hamiltonian, $\opr{H}=(2\mu)^{-1}\opr{p}^2$, i.e. $[\opr{H},\opr{T}]=i\hbar$. Equation (\ref{qtoa}) has been the subject of numerous investigations in $\Re$ and known to have unequal deficiency indices: $\opr{T}$ is not self-adjoint and lacks any self-adjoint extension in free unbounded space.\cite{freetime}. The non-self-adjointness of $\opr{T}$ has always been ascribed to the semiboundedness of the Hamiltonian in accordance with Pauli's theorem. But this is not necessarilly true. It is conceivable that even if $\opr{T}$ were self-adjoint and canonically conjugate to the Hamiltonian in some dense $\mcal{D}_c$, equations (\ref{com1}) and (\ref{com2}) remain unviolated as long as $\mcal{D}_c$ is not invariant under either $\opr{T}$ or $\opr{H}$. That is, $\opr{T}$ and $\opr{H}$ are canonically conjugate in some sense different from the position and momentum operators in an unbounded space.

Nevertheless equation (\ref{qtoa}) remains in the sorry state of non-self-adjointess in $\Re$. Can we explain this and in the process construct a self-adjoint version of $\opr{T}$? These we now attempt to do. In this letter we approach the problem with the additional assumption that the particle is known to be somewhere between two given points. That is the probability of finding the particle is zero outside and one in the entire length bounded by the two points. The particle is then essentially confined. Now let the length of the available spatial space be $2l$ and let the origin sit at the middle. If $p\neq 0$ and $\abs{q}<l$, classically the time of arrivat at the origin is still given by $T=-\mu qp^{-1}$; and $T$ remains canonically conjugate to the free Hamiltonian. Then equation (\ref{qtoa}) is still the totally symmetric quantized form of $T$ even when the particle is confined. We are then led to investigate equation (\ref{qtoa}) when the particle is confined.

At the level of equation (\ref{qtoa}), $\opr{T}$ is formal and its meaning is not precise until we define the domain and actions of the operators involved. We attach the Hilbert space $\mcal{H}=L^2[-l,l]$, the space of Lesbegue square integrable functions in the interval $[-l,l]$, to our system. We define the position, momentum, and Hamiltonian operators as follows
\begin{eqnarray}
\eqalign{\dom{q}=\mcal{H}\\ 
	\left(\opr{q}\psi\right)\!\!(q)=q\,\psi(q)\; \mbox{for all}\, \psi\in\dom{q}}\\
\eqalign{\dom{p_{\gamma}}=\left\{\phi\in\mcal{H}: \phi\, a.c., \phi'\in\mcal{H}, \phi(-l)=e^{-	2i\,\gamma}\phi(l),\,0\leq\gamma<1\right\}\\ 	
	\left(\opr{p}_{\gamma}\phi\right)(q)=\frac{\hbar}{i} \frac{d\phi}{dq}\; \mbox{for all}\, 	\phi\in\dom{p_{\gamma}}}\\ 
\eqalign{\dom{H_{\gamma}}=\left\{\varphi\in\dom{p_{\gamma}}:\, \varphi''\in\mcal{H},\, 	\varphi'(-l)=e^{-2i\,\gamma}\,\varphi'(l),\,0\leq\gamma<1\right\}\\ 	\left(\opr{H}_{\gamma}\varphi\right)\!\!(q)=-\frac{\hbar^2}{2\mu}\frac{d^2\varphi}{dq^2}\; 	\mbox{for all}\; \varphi\in\dom{H_{\gamma}}},
\end{eqnarray}

\noindent respectively. The different values of $\gamma$ correspond to different physics, e.g. what happens to a given wavepacket when it reaches one of the boundaries. As defined $\opr{q}$, $\opr{p}_{\gamma}$, and $\opr{H}_{\gamma}$ are densely defined, self-adjoint operators, with $\opr{q}$ and $\opr{p}_{\gamma}$ canonically conjugate in a dense subspace of $\dom{p_{\gamma}}$.  The Hamiltonian is so defined such that it is consistent with the interpretation that the energy is purely kinetic. The momentum and the Hamiltonian then commute and have the common set of eigenvectors $\left\{\phi_n^{(\gamma)}(q)=\frac{1}{\sqrt{2l}} \exp\!\left(i\,(\gamma+n \pi) \frac{q}{l}\right),\;n=0,\pm1,\pm2\cdots\right\}$; furthermore, both have pure point spectra.  According to current thinking, the Hamiltonian, being discrete, it can not be canonically conjugate to any self-adjoint operator. We will show the otherwise though.

Now we go back to equation (\ref{qtoa}), the formal operator $\opr{T}$. First for non-periodic boundary conditions, i.e. for $0<\gamma<1$. Since $\opr{q}$ appears in first power in (\ref{qtoa}), $\opr{T}$ is an operator if the inverse of $\opr{p}_{\gamma}$ exists.  The inverse exists if the range of $\opr{p}_{\gamma}$ is dense or its null space, $\nul{p_{\gamma}}$, is the trivial subspace. For non-periodic boundary conditions the constants do not belong to $\dom{p_{\gamma}}$, so that $\nul{p_{\gamma}}$ is the trivial subspace. Thus the inverse $\opr{p_{\gamma}^{-1}}$ exists. Because $\opr{p}_{\gamma}$ is unbounded and self-adjoint, $\opr{p_{\gamma}^{-1}}$ is bounded and likewise self-adjoint. Then it follows that for every $\gamma\in (0,1)$ $\opr{T}$ is a bounded, everywhere defined, \textsl{self-adjoint} operator! For a given $\gamma$ we identify $\opr{T}$ with the operator $\opr{T}_{\gamma}=\mu(\opr{q}\opr{p_{\gamma}^{-1}}+\opr{p_{\gamma}^{-1}}\opr{q})2^{-1}$ derived from the formal operator (\ref{qtoa}) by replacing $\opr{p}$ with $\opr{p}_{\gamma}$. We shall refer to $\opr{T}_{\gamma}$ as the confined, non-periodic time of arrival operator for a given $\gamma\in(0,1)$.

Now we seek an explicit coordinate representation of $\opr{T}_{\gamma}$. We proceed by noting that $\opr{p_{\gamma}^{-1}}$ has the representation	$(\opr{p_{\gamma}^{-1}}\varphi)\!(q)=\frac{l}{\hbar}\!\sum_{n=-		\infty}^{\infty}(\phi_n^{(\gamma)},\varphi)(\gamma+n\pi)^{-1} 		\phi_n^{(\gamma)}(q)$ for all $\varphi(q)\in\mcal{H}$.
This leads to the Fredholm integral operator representation of $\opr{T_{\gamma}}$,
\begin{equation} \label{fredholm}
	\left(\opr{T}_{\gamma}\varphi\right)\!(q)=\int_{-l}^{l}T_{\gamma}(q,q')\,\varphi(q')\,dq' 		\;\;\;\mbox{for all}\;\;\;\varphi(q)\in\mcal{H},
\end{equation}

\noindent where the kernel is given by
\begin{eqnarray} \label{repre}
	T_{\gamma}(q,q')&=&-\frac{\mu}{4\hbar}(q+q') \!\sum_{n=-\infty}^{\infty} 		\frac{\phi_n^{(\gamma)}(q)\,\phi_n^{*(\gamma)}(q')}{\gamma+n\pi} \nonumber \\
			    &=&-\frac{\mu}{4\hbar\,\sin\!\!\gamma} (q+q')\left(e^{i\,\gamma} 				\,\mbox{H}(q-q')+e^{-i\,\gamma}\, \mbox{H}(q'-q) \right)
\end{eqnarray}

\noindent The kernel $T_{\gamma}(q,q')$ is both symmetric and bounded, i.e. $T_{\gamma}(q,q')=T_{\gamma}^{*}(q',q)$ and $\int_{[-l\times l]} \abs{T_{\gamma}(q,q')}^2\,dq\,dq'<\infty$, respectively. This reaffirms the self-adjointness of $\opr{T_{\gamma}}$. 

Now it is straightforward to show that $\opr{H}_{\gamma}$ and $\opr{T}_{\gamma}$  for a given $0<\gamma<1$ are canonically conjugate in the following sense:
	$\left((\opr{H_{\gamma}T_{\gamma}}-\opr{T_{\gamma}H_{\gamma}})\varphi\right)\!\!(q) =i\,\hbar\,\varphi(q)$  for all $\varphi(q)\in \mcal{D}_c^{(\gamma)}$ where $\mcal{D}_c^{(\gamma)}=\left\{\varphi(q)\in\dom{H_{\gamma}}:\int_{-l}^{l} \varphi(q)\,dq=0,\, \varphi(\partial)=0,\, \varphi'(\partial)=0 \right\}$. The subspace ${D}_c^{(\gamma)}$ is explicitly given by 
\begin{equation} \label{domain}
\fl \mcal{D}_c^{(\gamma)}= \left\{\varphi(q)=\sum_{n=1}^{\infty}\varphi_n\,\phi_n^{(\gamma)}(q),\, \sum_{n=1}^{\infty} n^2\abs{\varphi_n}^2<\infty,\, \sum_{n=1}^{\infty} \frac{(-1)^n 				n\,\varphi_n}{\sqrt{\gamma^2+\pi^2 n^2}}=0\right\}
\end{equation}

\noindent where $\phi_n^{(\gamma)}=(l\gamma^2+l\pi^2 n^2)^{-\frac{1}{2}}\left(i\,\gamma \sin\frac{n\pi q}{l}+\, n\pi \cos\frac{n\pi q}{l}\right)\exp(\frac{i\gamma q}{l})$. Moreover $\mcal{D}_c^{(\gamma)}$ is orthogonal to the subspace spanned by the states $\phi_0^{(\gamma)\perp}=(2l)^{-\frac{1}{2}}\exp(\frac{i\gamma q}{l})$, $\phi_n^{(\gamma)\perp}=(l\gamma^2+l\pi^2 n^2)^{-\frac{1}{2}}\left(i\,\gamma \cos\!\!\frac{n\pi q}{l}+\,n\,\pi \sin\!\!\frac{n\pi q}{l}\right)  \exp\left(\frac{i\gamma q}{l}\right)$, $n=1,2\cdots$. Thus  $\mcal{D}_c^{(\gamma)}$ is closed. 

For periodic boundary conditions, $\gamma=0$, we face a different problem. Now the range of the momentum operator $\opr{p_0}$ is no longer dense, because its null space, $\nul{p_0}$, consists of the non-trivial one dimensional subspace spanned by the state $\phi_0^{(0)}(q)$, the state of vanishing momentum. The inverse of $\opr{p_0}$ then does not exist and equation (\ref{qtoa}) is meaningless. This is reflected by the ill defined limit of the kernel (\ref{repre}) in the limit as $\gamma\rightarrow 0$. Evidently the pathology arises from $\nul{p_0}$. But this subspace is not relevant to the problem because the question of when a given particle arrives makes sense only when the particle is in motion, otherwise it does not get anywhere (see also ref.\cite{grot}). We then expect that, with $l$ fixed, the non-periodic kernel (\ref{repre}) has a finite part corresponding to the non-vanishing momentum components in the limit as $\gamma\rightarrow 0$. Indeed this is the case. The finite part can be extracted by removing the divergent contribution of the vanishing momentum eigenvalue to give
\begin{equation} \label{periodic}
	T_{0}(q,q')=i\frac{\mu}{4\hbar}(q+q')\mbox{sgn}(q-q')-i\frac{\mu}{4\hbar l}\left(q^2 -q'^2\right)
\end{equation}

\noindent Equation (\ref{periodic}) is also symmetric and bounded. This means that the finite periodic limit of (\ref{repre}) generates a self-adjoint integral operator, $\opr{T_0}$, whose kernel is given by equation (\ref{periodic}). $\opr{T}_0$ can be interpreted as the time of arrival operator for periodic boundary conditions for states with non-vanishing momentum. We shall refer to this as the confined, periodic time of arrival operator.  It can be shown that $\opr{H_0}$ and $\opr{T_0}$ are also canonically conjugate; that is, $\left(\left(\opr{H_0 T_0}-\opr{T_0 H_0}\right)\varphi\right)\!\!(q)=i\hbar\,\varphi(q)$ for all $\varphi\in\mcal{D}_c = \left\{\varphi\in\dom{H_0}:\,\varphi^{(k)}(\partial)=0,\int_{-l}^{l} q^{k}\,\varphi(q)\,dq=0,\, k=0,1\right\}$. Moreover $\mcal{D}_c=\mcal{D}_c^{(0)}$ (refer to equation (\ref{domain})) so that $\phi_n^{(0)}$ spans $\mcal{D}_c$. Also $\mcal{D}_c$ is orthogonal to the subspace spanned by the states $\phi_n^{(0)\perp}$. Thus $\mcal{D}_c$ is likewise closed. 

We point out that the pair $(\opr{T}_{\gamma},\opr{H}_{\gamma})$  for every $\gamma\in[0,1)$ satisfy the canonical commutation relation without violating equations (\ref{com1}) and (\ref{com2}) because $\mcal{D}_c^{(\gamma)}$ is closed: equations (\ref{com1}) and (\ref{com2}) imply that the generators of $U_{\alpha}$ and $V_{\beta}$ are canonically conjugate in a dense subspace $\mcal{D}_c$ which is invariant under both generators. In our terminology, the pair $(\opr{H}_{\gamma},\opr{T}_{\gamma})$ are conjugate in a sense different from the generators of $U_{\alpha}$ and $V_{\beta}$. The sense that the pair $(\opr{T}_{\gamma},\opr{H}_{\gamma})$ is canonical can be further appreciated by noting that $\mcal{D}_c^{(\gamma)}$ carries information on the states with zero time of arrival expectation values. It can be shown that the states spanning $\mcal{D}_c^{(\gamma)}$ and $\mcal{D}_c^{(\gamma)\perp}$---the states $\{\phi_n^{(\gamma)},\phi_n^{(\gamma)\perp}\}$---have zero time of arrival expectation values. Morevover, $\mcal{D}_c^{\gamma}$ identifies the states for which the the TOA-energy uncertainty relation holds, i.e. $\Delta T_{\gamma}(\varphi) \Delta E_{\gamma}(\varphi)\ge \hbar\, 2^{-1}$ for all $\varphi\in\mcal{D}_c^{(\gamma)}$. $\opr{T}_{\gamma}$ being bounded,  the uncertainty relation does not hold outside $\mcal{D}_c^{\gamma}$, i.e. $\Delta T_{\gamma}(\varphi) \Delta E_{\gamma}(\varphi)\ge 0$ for all $\varphi\in\mcal{H}\setminus\mcal{D}_c^{(\gamma)}$.

Can we physically explain our above results and relate to the non-self-adjointness of equation (\ref{qtoa}) in unbounded space?  When we ask ''when will the particle arrive?'' we suppose that the particle was prepared in a state that will {\sl arrive}; otherwise, we don't ask when. Classically, for a free particle, the arrival is assured by imposing that the position is finite and it has non-vanishing momentum. When either of these is not satisfied, we are dommed to wait indefinitely. Now how do we translate these conditions in quantum mechanics? We may impose that the physical Hilbert space consists only of states with bounded supports in spatial space and allows equation (\ref{qtoa}) to be well defined in the neighborhood of vanishing momentum. For the confined particle for non-periodic boundary conditions, both of these are satisfied. And equation (\ref{qtoa}) is bounded and self-adjoint.  For periodic boundary conditions, the first is satisfied yet (\ref{qtoa}) is ill defined. And this is because the spectrum of $\opr{p}_0$ includes the zero eigenvalue, a non-arrival feature. A systematic removal of this non-arrival feature leads to a bounded and self-adjoint TOA. And the boundedness of these confined TOA operators assures that the particle will arrive within a finite interval of time. For the free particle in unbounded space, the topology induced by the $L^2$ norm accomodates states whose tails extend to infinity and yields states in the domain of the adjoint of (\ref{qtoa}) that equation (\ref{qtoa}) is otherwise undefined. It then appears that the non-self-adjointness or not being an operator of equation (\ref{qtoa}) at all is a consequence of the residual non-arrival features of $T$ upon naive quantization. And removal of these non-arrival features has led us to a class of confined self-adjoint and canonical TOA-operators. In a seperate publication we shall deal with the $\opr{T}_{\gamma}$'s at a deeper level, e.g. their spectral properties and measurements, and explore the relationship between our results and the POVM program \cite{egus}.

\end{document}